\documentclass{ws-procs9x6}
\def\half{\frac{1}{2}}
\begin{document}
\title{GEOMETRIC SPECTRAL INVERSION}
\author{Richard L. Hall}
\address{\address{Department of Mathematics and Statistics, Concordia University,
1455 de Maisonneuve Boulevard West, Montreal,
Quebec, Canada H3G 1M8}
$^*$E-mail: rhall@mathstat.concordia.ca\\
www.mathstat.concordia.ca/faculty/rhall}
\begin{abstract}
A discrete eigenvalue $E_n$ of a Schr\"odinger operator $H = -\Delta + vf(r)$ is given, as a function $F_n(v)$ of the coupling parameter $v\ge v_c.$ It is shown how the potential shape $f(x)$ can be reconstructed from $F_n(v).$
A constructive inversion algorithm and a functional inversion sequence are both discussed.
\end{abstract}
\bodymatter
\section{Introduction}
We suppose that a discrete eigenvalue $E_n = F_n(v)$ of the Schr\"odinger Hamiltonian
$$H = -\Delta + vf(x)$$
is known for all sufficiently large values of the coupling parameter $v \ge v_c$ and we use this data to reconstruct the potential shape $f(x).$  The usual `forward' problem would be: given the potential (shape) $f(x),$ find the corresponding energy trajectories $F_n(v).$ 
For example, if the potential shape is $f(x) = - {\rm sech}^2(x), $ then the eigenvalues as functions of the coupling $v$  are given \cite{flug} by the formula
\[
F_n(v) = -\left[\left(v+{1 \over 4}\right)^{\half} - \left(n+\half\right)\right]^{2},\quad n = 0,1,2,\dots,\quad v \ge n(n+1).\eqno{(1.1)}
\]
These energy graphs are illustrated in Fig.(1).

\begin{figure}
\psfig{file=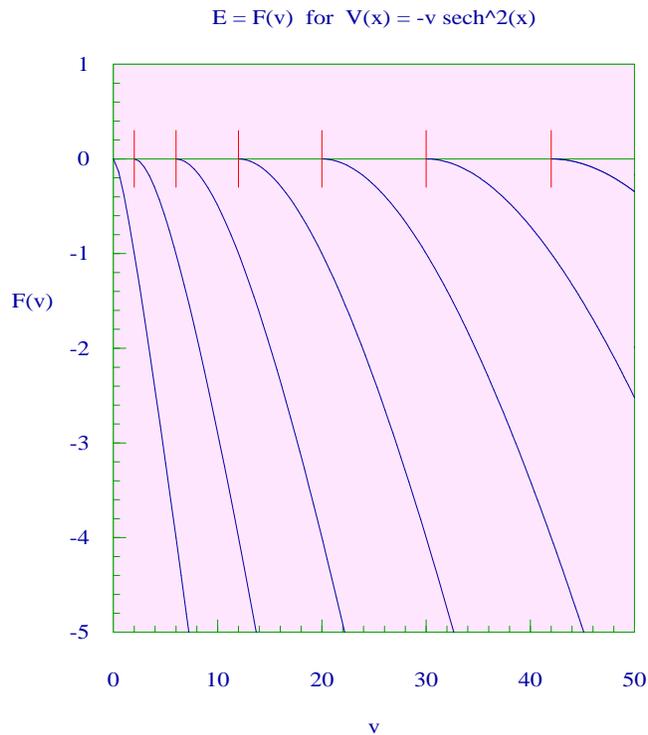,width=4.5in}
\caption{Discrete eigenvalues of $H = -\Delta -v~ {\rm sech}^2(x)$
 as functions of the coupling $v > 0.$}
\label{sechfv:fig1}
\end{figure}

The problem we now consider here is the inverse of this, namely $F \rightarrow f.$  We call this problem `geometric spectral  inversion'.  It must at once be distinguished both from inverse scattering theory
 \cite{chad,newt,zak,toda,eilen} and, more specifically, from the `inverse problem in the coupling constant' discussed, for example, by Chadan {\it et al} \cite{chad,chada,chadb,chadc,chadd}. In this latter problem, the discrete part of the `input data' is a set $\{v_{i}\}$ of values of the coupling constant that all yield the identical energy eigenvalue $E.$ The index $i$ might typically represent the number of nodes in the corresponding eigenfunction.  In contrast, for the problem discussed in the present paper, $i$ is kept fixed and the input data is the graph $(F(v),v),$ where the coupling parameter has any value $v > v_c,$ and $v_c$ is the critical value of $v$ for the support of a discrete eigenvalue with $i$ nodes.  Geometric spectral inversion has been discussed  in a series of earlier papers \cite{inv1,inv2,inv3,inv4,inv5,inv6}. Here we shall mainly discuss the bottom of the spectrum $i = 0$. However, on the basis of results we have obtained for the inversion IWKB of the WKB approximation \cite{inv3}, there is good reason to expect that inversion is possible starting from any discrete eigenvalue trajectory $F_{n}(v),$ $n > 0.$  In fact, perhaps not surprisingly, IWKB yields better results starting from higher trajectories; moreover, they become asymptotically exact as the eigenvalue index is increased without limit.

After recalling some basic results  in the remainder of this introduction, in sections~2-4 we discuss a constructive algorithm for generating $f(x)$ from $F(v).$ In section-5 we apply this method to some examples. In sections~6 and 7, we briefly describe an established geometric spectral theory \cite{env1,env2,env3,env4,env5,env6} that enables us in section~8 to effect `functional inversion' \cite{inv6}.  This  latter method allows us to start with a seed potential $f^{[0]}(x)$ and from this reconstruct $f(x)$ by means of a sequence of functional operations. This latter method is applied to some examples in sections~9 and 10.

By making suitable assumptions concerning the class of potential shapes, general theoretical progress has already been made with this inversion problem \cite{inv1,inv2}.  The assumptions that we retain for most of the present paper are that $f(x)$ is symmetric, monotone increasing for $x >0,$ and bounded below: consequently the minimum value is $f(0).$  We assume that our spectral data, the energy trajectory $F(v),$ derives from a potential shape $f(x)$ with these features.  We have discussed \cite{ref1}  how two potential shapes $f_1$ and $f_2$ can cross over and still preserve spectral ordering $F_1 < F_2.$  It is known \cite{inv2} that lowest point $f(0)$ of $f$ is given by the limit
$$f(0) = {\lim_{v\rightarrow\infty}} {F(v) \over v}.\eqno{(1.2)}$$
We have proved \cite{inv1} that a potential shape $f$ has a finite flat portion ($f'(x) = 0$) in its graph starting at $x = 0$ if and only if the mean kinetic energy is bounded.  That is to say, $s = F(v) - vF'(v) \le K,$ for some positive number $K.$  More specifically, the size $b$ of this patch can be estimated from $F$ by means of the inequality:
$$s \leq K \quad \Rightarrow\quad f(x) = f(0),\quad |x| \leq b, \quad {\rm and} \quad b = {\pi \over 2} K^{-{1 \over 2}}.\eqno{(1.3)}$$
The monotonicity of the potential, which allows us to prove results like this, also yields the\hfil\break 
\noindent{\bf Concentration Lemma} \cite{inv1}
$$q(v) = \int_{-a}^{a}\psi^2(x,v)dx > {{f(a) - F'(v)} \over {f(a) - f(0)}} \quad \rightarrow \quad 1, \quad{\rm as}~ v \rightarrow \infty,\eqno{(1.4)}$$
where $\psi(x,v)$ is the normalized eigenfunction satisfying $H\psi = F(v)\psi.$  More importantly, perhaps, if $F(v)$ derives from a symmetric monotone potential shape $f$ which is bounded below, then $f$ is {\it uniquely} determined \cite{inv2}. The significance of this result can be appreciated more clearly upon consideration of an example. Suppose the bottom of the spectrum of $H$ is given by $F(v) = \sqrt{v},$ what is $f(x)$?  It is well known, of course, that $f(x) = x^{2} \rightarrow F_{o}(v) = \sqrt{v};$ but are there any other potential pre-images of this spectral function $F_o(v)$? Are scaling arguments reversible? A possible source of initial disquiet for anyone who ponders such questions is the uncountable number of (unsymmetric) perturbations \cite{raab} of the harmonic oscillator $f(x) = x^2$ there are, all of which have the identical spectrum $E_n = (2n+1), n = 0,1,2,\dots,$ to that of the unperturbed oscillator (with coupling $v=1$).

If, in addition to symmetry and monotonicity, we also assume that a potential shape $f_{a}(x)$ vanishes at infinity and that $f_{a}(x)$ has area,  then a given trajectory function $F_{a}(v)$ corresponding to $f_{a}(x)$ can be `scaled' \cite{inv2} to a standard form in which the new function $F(v) = \alpha F_{a}(\beta v)$ corresponds to a potential shape $f(x)$ with area $-2$ and minimum value $f(0) = -1.$  Thus square-well potentials, which of course are completely determined by depth and area, are immediately invertible; moreover it is known that,  amongst all standard potentials, the square-well is `extremal' for it has the lowest possible energy trajectory.  In Ref.\cite{inv2} an approximate variational inversion method is developed; it is also demonstrated constructively that all separable potentials are invertible.  However, these results and additional constraints are not used in the present paper.  When a potential has area $2A$, we first assumed, during our early attempts at numerical inversion, that it would be very useful to determine $A$ from $F(v)$ and then appropriately constrain the inversion process. However, the area constraint did not turn out to be helpful.  Thus the numerical method we have established for constructing $f(x)$ from $F(v)$ does not depend on use of this constraint, and is therefore not limited to the reconstruction of potentials which vanish at infinity and have area.
\section{Constructive inversion} 
Much of numerical analysis assumes that errors arising from arithmetic computations or from the computation of elementary functions is negligibly small.  The errors usually studied in depth are those that arise from the discrete representation of continuous objects such as functions, or from operations on them, such as derivatives or integrals.  In this paper we shall take this separation of numerical problems to a higher level.  We shall assume that we have a numerical method for solving the eigenvalue problem in the {\it forward} direction $f(x) \rightarrow F(v)$ that is reliable and may be considered for our purposes to be essentially error free.  Our main emphasis will be on the design of an effective algorithm for the inverse problem {\it assuming} that the forward problem is numerically soluble.  The forward problem is essential to our methods because we shall need to know not only the given exact energy trajectory $F(v)$ but also, at each stage of the reconstruction, what eigenvalue a partly reconstructed potential generates. This line of thought immediately indicates that we shall also need a way of temporarily extrapolating a partly reconstructed potential to all $x.$

Our constructive inversion algorithm hinges on the assumed symmetry and monotonicity of $f(x).$ This allows us to start the reconstruction of $f(x)$ at $x = 0,$ and sequentially increase $x$. In Section (2) it is shown how numerical estimates can be made for the shape of the potential near $x = 0,$ that is for $x < b,$ where $b$ is a parameter of the algorithm. In Section (3) we explore the implications of the potential's monotonicity for the `tail' of the wave function.   In Section (4) we establish a numerical representation for the form of the unknown potential for $x > b$ and construct our inversion algorithm.  In Section (5) the algorithm is applied to three test problems.       
 \subsection{The reconstruction of $f(x)$ near $x = 0.$}
Since the energy trajectory $F(v)$ which we are given is assumed to arise from a symmetric monotone potential, and since the spectrum generated by the potential is invariant under shifts along the $x$-axis, we may assume without loss of generality that the minimum value of the potential occurs at $x = 0.$  We now investigate the behaviour of $F(v),$ either analytically or numerically, for large values of $v.$  The purpose is to establish a value for the starting point $x = b > 0$ of our inversion algorithm and the shape of the potential in the interval $x \in [0,b].$  First of all, the minimum value $f(0)$ of the potential is provided by the limit (1.2).  Now, if the mean kinetic energy $s = (\psi, -\Delta \psi) = F(v) - vF'(v)$ is found to be bounded above by a positive number $K,$ then we know \cite{inv1} that the potential shape $f(x)$ satisfies $f(x) = f(0),\quad x \in [0,b],$ where b is given by (1.3).  In this case we have a value for $b$ and also the shape $f(x)$ inside the interval $[0,b].$

If the mean potential energy $s$ is (or appears numerically to be) unbounded, then we adopt another strategy: we model $f(x)$ as a shifted power potential near $x = 0.$  Since we never know $f(x)$ {\it exactly,} we shall need another symbol for the approximation we are currently using for $f(x).$   We choose this to be $g(x)$ and we suppose that the bottom of the spectrum of $-\Delta + vg(x)$ is given by $G(v).$  The goal is to adjust $g(x)$ until $G(v)$ is close to the given $F(v).$  Thus we write
$$f(x) \approx g(x) = f(0) + Ax^{q},\quad x \in [0,b].\eqno{(2.1)}$$
Therefore we have three positive parameters to determine, $b,\ A,$ and $q.$  We first suppose  that $g(x)$ has the form (2.1) for {\it all} $x \geq 0.$  We now choose a `large' value $v_{1}$ of $v.$  This is related to the later choice of $b$ by a bootstrap argument: the idea is that we choose $v_{1}$ so large that the turning point determined by 
$$\psi_{xx}(x,v_1)/\psi(x,v_1) = v_{1}f(x) - F(v) = 0\eqno{(2.2)}$$
is equal to $b.$  The concentration lemma guarantees that this is possible.  By scaling arguments we have 
$$G(v) = f(0)v + E(q)(vA)^{2 \over {2+q}},\eqno{(2.3)}$$
where $E(q)$ is the bottom of the spectrum of the pure-power Hamiltonian $-\Delta + |x|^{q}.$  We now `fit' $G(v)$ to $F(v)$ by the equations $G(v_{1}) = F(v_{1})$ and $G(2v_{1}) = F(2v_{1})$  which yield the estimate for $q$ given by
$$\eta = {2 \over {2 + q}} = {{\log(F(2v_{1})-2v_{1}f(0)) - \log(F(v_{1})-v_{1}f(0))} \over {\log(2)}}.\eqno{(2.4)}$$
Thus $A$ is given by
$$A = \left((F(v_{1}) - v_{1}f(0))/E(q)\right)^{1 \over \eta}/v_{1}.\eqno{(2.5)}$$
We choose $b$ to be equal to the turning point corresponding to the {\it model} potential $g(x)$ with the smaller value of $v,$ that is to say so that $f(0) + Ab^{q} = F(v_{1})/v_{1},$ or
$$b = \left({{F(v_{1})-v_{1}f(0)} \over {Av_{1}}}\right)^{1 \over q}.\eqno{(2.5)}$$
Thus we have determined the three parameters which define the potential model $g(x)$ for $x \in [-b,b].$  
    \section{The tail of the wavefunction} 
Let us suppose that the ground-state wave function is $\psi(x,v).$ Thus the turning point $\psi_{xx}(x,v) = 0$ occurs for a given $v$ when 
$$x = x_{t}(v) = f^{-1}(R(v)),\quad R(v) = \left({{F(v)} \over v}\right).\eqno{(3.1)}$$
The concentration lemma (1.4) quantifies the tendency of the wave function to become, as the coupling $v$ is increased, progressively more concentrated on the patch $[-c,c],$ where $x = c$ is the point (perhaps zero) where $f(x)$ first starts to increase.  This allows us to think in terms of the wave function having a `tail'.  We think of a symmetric potential as having been determined from $x = 0$ up to the current point $x.$  The question we now ask is: what value of $v$ should we use to determine how $f(x),$ or, more particularly, our {\it approximation} $g(x)$ for $f(x),$ continues beyond the current point.  We have found that a good choice is to choose $v$ so that the turning point $x_{t}(v) = x/2,$ or some other similar fixed fraction $\sigma < 1$ of the current $x$ value. The algorithm seems to be insensitive to this choice.  Since $g(x)$ has been constructed up to the current point, and $F(v)$ is known, the value of $v$ required follows by inverting (3.1).  It has been proved \cite{inv2} that $R(v)$ is monotone and therefore invertible.  Hence we have the following general recipe for $v:$ 
$$v = R^{-1}(g(\sigma x)),\quad \sigma = {1 \over 2}.\eqno{(3.2)}$$
Since we can only determine Schr\"odinger eigenvalues of $H = -\Delta + vg(x)$ if the potential is defined for all $x,$ we must have a policy about temporarily extending $g(x).$   We have tried many possibilities and found the simplest and most effective method is to extend $g(x)$ in a straight line, with slope to be determined.
\begin{figure}
\psfig{file=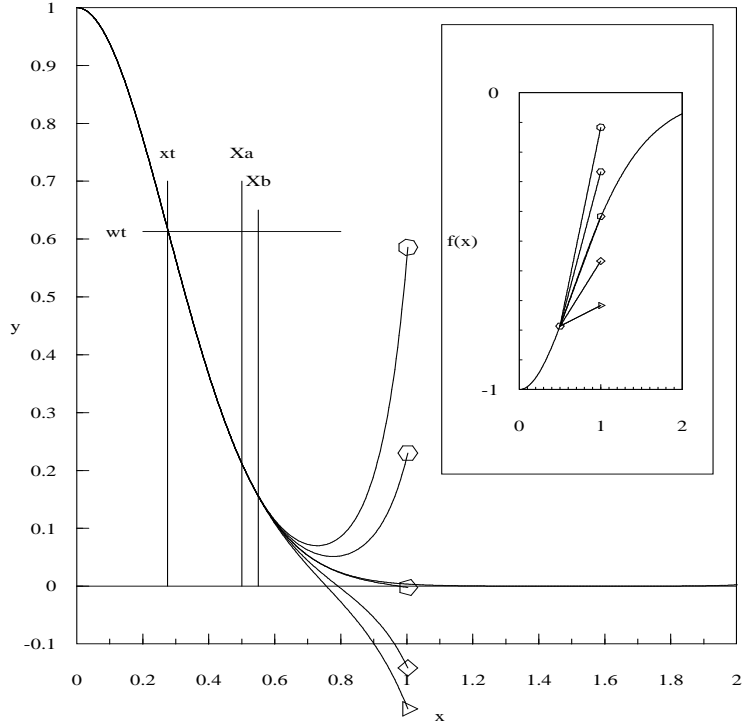,width=4in,height=4in}
\caption{We illustrate the ideas discussed in the text for the case of the sech-squared potential.  The inset graph shows the sech-squared potential perturbed from $x = x_{a}$ by five straight line extensions; meanwhile the main graph shows the corresponding set of five wave functions which agree for $0 \leq x \leq x_{a}$ and then continue with different `tails' dictated by the corresponding potential extensions.  The value of the coupling $v$ is the value that makes the turning point of the wave function occur at $x = x_{a}/2.$  This figure illustrates the sort of graphical study that has lead to the algorithm described in this paper.}
\label{fig2}
\end{figure}
      \section{Constructive inversion algorithm}
We must first define the `current' approximation $g(x)$ for the potential $f(x)$ sought. For values of $x$ less than $b,$ $g(x)$ is defined either as the horizontal line $f(x) = f(0)$ or as the shifted power potential (2.1).  For values of $x$ greater than $b,$ the $x$-axis is divided into steps of length $h.$ Thus the `current' value of $x$ would be of the form $x = x_{k} = b + kh,$ where $k$ is a positive integer.  The idea is that $g(x_{k})$ is determined sequentially and $g(x)$ is interpolated linearly between the $x_{k}$ points.  We suppose that $\lbrace g(x_{k})\rbrace$ have already been determined up to $k$ and we need to find $y = g(x_{k+1}).$  For $x \geq x_{k}$ we let
$$g(x) = g(x_{k}) + (y - g(x_{k})) {{x - x_{k}} \over h}.\eqno{(4.1)}$$
If, from a study of $F(v),$ the underlying potential $f(x)$ has been shown \cite{inv2} to be bounded above, it is convenient to rescale $F(v)$ so that it corresponds to a potential shape $f(x)$ which vanishes at infinity.  In this case  it is slightly more efficient to modify (4.1) so that for large $x$ the straight-line extrapolation of $g(x)$ is `cut' to zero instead of becoming positive.  In either case we now have for the current point $x_{k}$ an approximate potential $g(x)$ parameterized by the `next' value $y = g(x_{k+1}).$  The task of the inversion algorithm is simply to choose this value of $y.$

Let us suppose that, for given values of $k$ and $y,$ the bottom of the spectrum of $H = -\Delta + vg(x)$ is given by $G(v,k,y),$ then the inversion algorithm may be stated in the following succinct form in which $\sigma < 1$ is a fixed parameter.  Find $y$ such that 
$$vg(\sigma x_{k}) = F(v) = G(v,k,y);\quad {\rm then}\quad g(x_{k+1}) = y.\eqno{(4.2)}$$
The value of $v$ is first chosen so that the turning point of the wave function generated by $g$ occurs at $\sigma x_{k};$ after this, the value of $y$ is chosen so that $G$ `fits' $F$ for this value of $v.$  The value of the parameter $\sigma$ chosen for the examples discussed in section~(5) below is $\sigma = {1 \over 2}.$   The idea behind this choice can best be understood from a study of Figure~(1): the value of the coupling $v$ must be such that the current value of $x$ for which $y$ is sought is in the `tail' of the corresponding wave function; that is to say, the turning point $\sigma x$ should be before $x,$ but not too far away.  Fortunately the inversion algorithm seems to be insensitive to the choice of $\sigma.$
      \section{Some examples}
The first example we consider is the unbounded potential whose shape $f(x)$ and corresponding exact energy trajectory $F(v)$ are given by the $\{f,F\}$ pair
$$f(x) = -1 + |x|^{3 \over 2}\quad \longleftrightarrow \quad F(v) = -v + E(3/2)v^{4 \over 7},\eqno{(5.1)}$$
where $E(3/2)$ is the bottom of the spectrum of $H = -\Delta + |x|^{3 \over 2}$ and has the approximate value $E(3/2) \approx 1.001184.$  Applying the inversion algorithm to $F(v)$ we obtain the reconstructed potential shown in Figure~(3).
\begin{figure}
\psfig{file=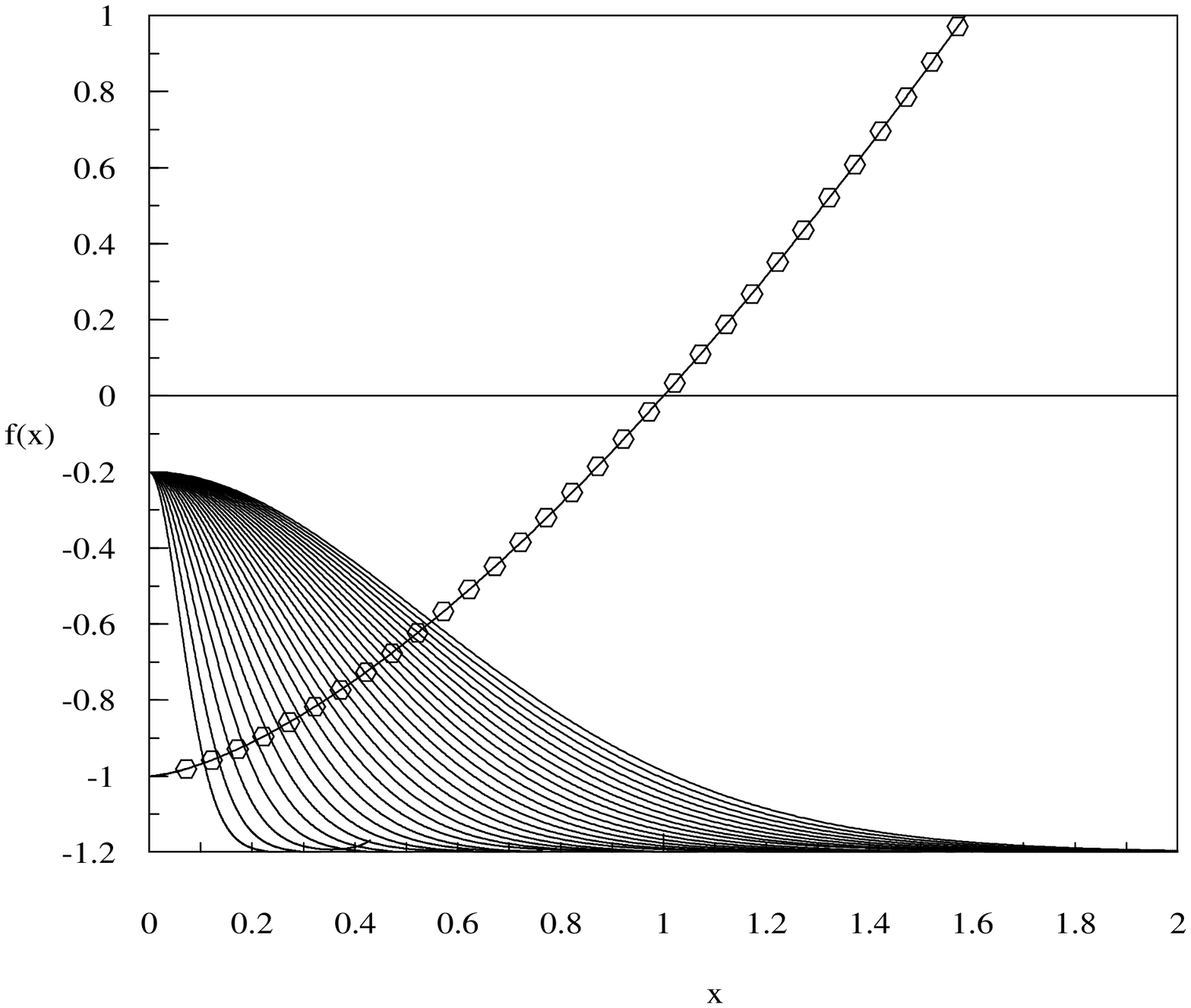,width=4in,height=3in}
\caption{Constructive inversion of the energy trajectory $F(v)$ for the shifted
 power potential $f(x) = -1 + |x|^{3 \over 2}.$  For $x \leq b = 0.072,$ the algorithm correctly
 generates the model $f(x);$ for larger values of $x,$ in steps of size $h = 0.05,$ the hexagons
 indicate the reconstructed values for the potential $f(x),$ shown exactly as a smooth curve. 
The unnormalized wave functions are also shown.}
\label{fig3}
\end{figure}
We first set $v_1 = 10^4$ and find that the initial shape is determined (as described in Section~(2)) to be $-1 + x^{1.5}$ for $x < b = 0.072.$ For larger values of $x$ the step size is chosen to be $h = 0.05$ and $40$ iterations are performed by the inversion algorithm.  The results are plotted as hexagons on top of the exact potential shape shown as a smooth curve. 

The following two examples are bounded potentials both having large-$x$ limit zero, lowest point $f(0) = -1,$ and `area' $-2.$ The exponential potential \cite{flug, mass} has the $\{f,F\}$ pair
$$f(x) = -e^{-|x|}\quad \longleftrightarrow \quad J'_{2|E|^{1 \over 2}}(2v^{1 \over 2}) = 0 \quad \equiv\quad E = F(v),\eqno{(5.2)}$$
where $J'_{\nu}(x)$ is the derivative of the Bessel function of the first kind of order $\nu.$ \medskip

\noindent For the sech-squared potential \cite{flug} we have 
$$f(x) = -{\rm sech}^{2}(x)\quad \longleftrightarrow \quad F_0(v) = -\left[\left(v+{1 \over 4}\right)^{\half} - \frac{1}{2}\right]^{2}.\eqno{(5.3)}$$ 
In Figure~(4) the two energy trajectories are plotted.
\begin{figure}
\psfig{file=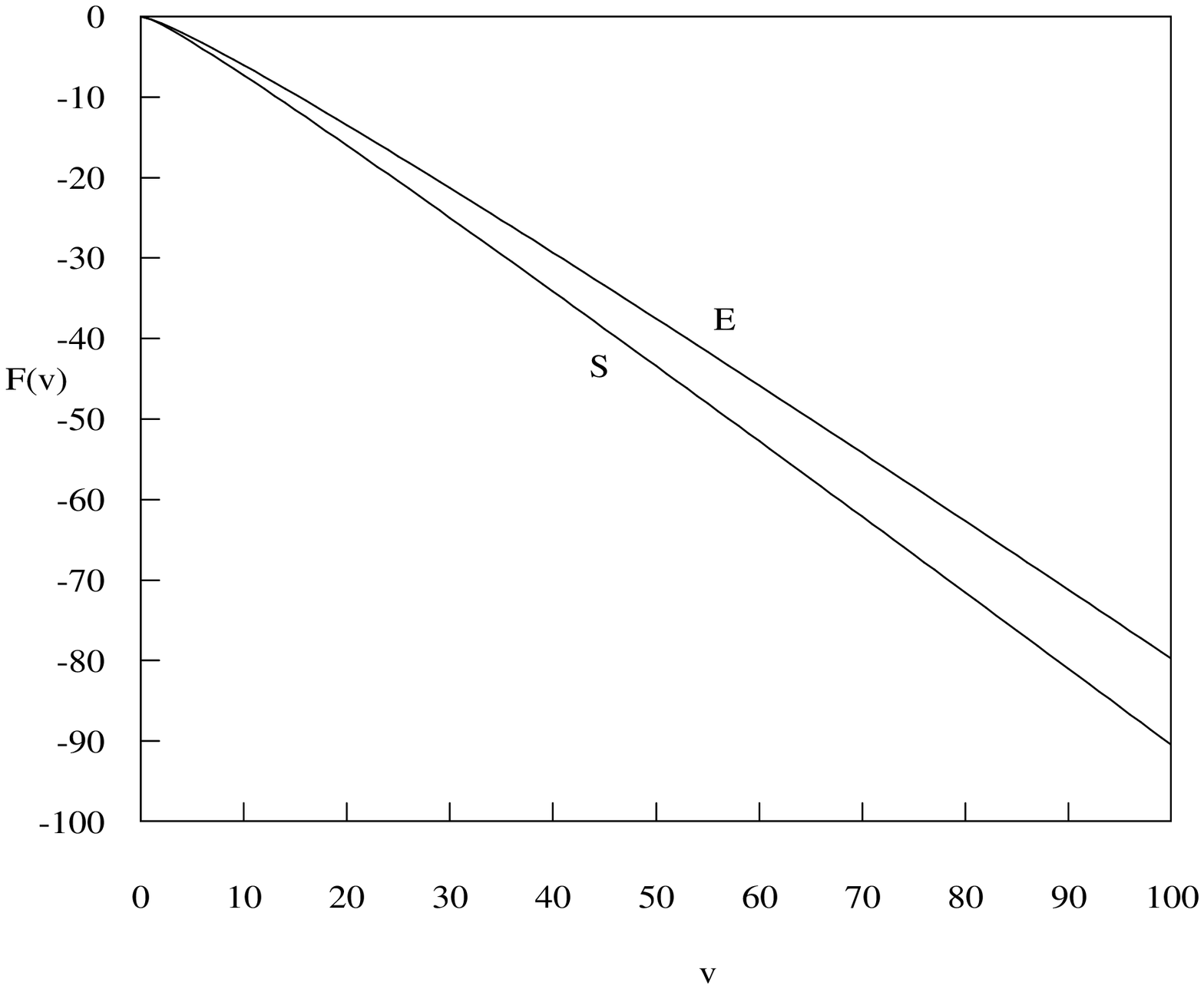,width=4in,height=3in}
\caption{The ground-state energy trajectories $F(v)$ for the exponential potential
 (E) and the sech-squared potential (S). For small $v,$ $F(v) \approx -v^{2};
$ for large $v,$ $\lim_{v \rightarrow \infty}\left(F(v)/v\right) = -1.$  The shapes of the underlying
 potentials are buried in the details of $F(v)$ for intermediate values of $v.$}
\label{fig4}
\end{figure}
Since the two {\it potentials} have lowest value $-1$ and `area' $-2$ it follows \cite{inv2} that the corresponding trajectories both have the form $F(v) \approx -v^{2}$ for small $v$ and they both satisfy the large-$v$ limit $\lim_{v \rightarrow \infty}\left(F(v)/v\right) = -1.$  Thus the differences between the potential shapes is somehow encoded in the fine differences between these two similar energy curves for intermediate values of $v:$ it is the task of our inversion theory to decode this information and reveal the underlying potential shape.  If we apply the inversion algorithm to these two problems we obtain the results shown in Figures~(5) and (6). 
\begin{figure}
\psfig{file=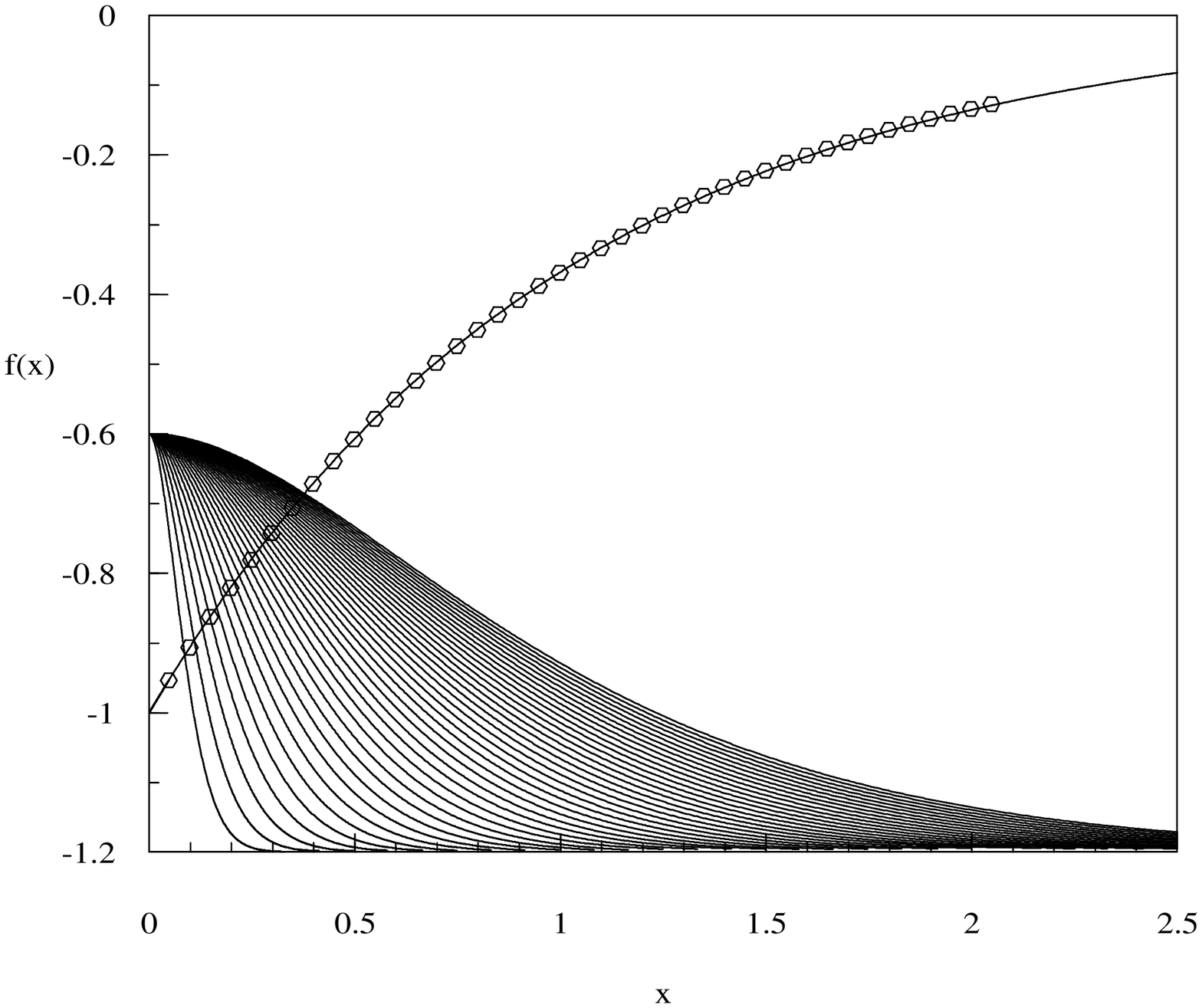,width=4in,height=3in}
\caption{Constructive inversion of the energy trajectory $F(v)$ for the
 exponential potential $f(x) = -\exp(x).$  For $x \leq b = 0.048,$ the algorithm correctly
 generates the model $f(x) = -1 + |x|;$ for larger values of $x,$ in steps of size $h = 0.05,$
 the hexagons indicate the reconstructed values for the potential $f(x),$ shown exactly as a
 smooth curve. The unnormalized wave functions are also shown.}
\label{fig5}
\end{figure}
\begin{figure}
\psfig{file=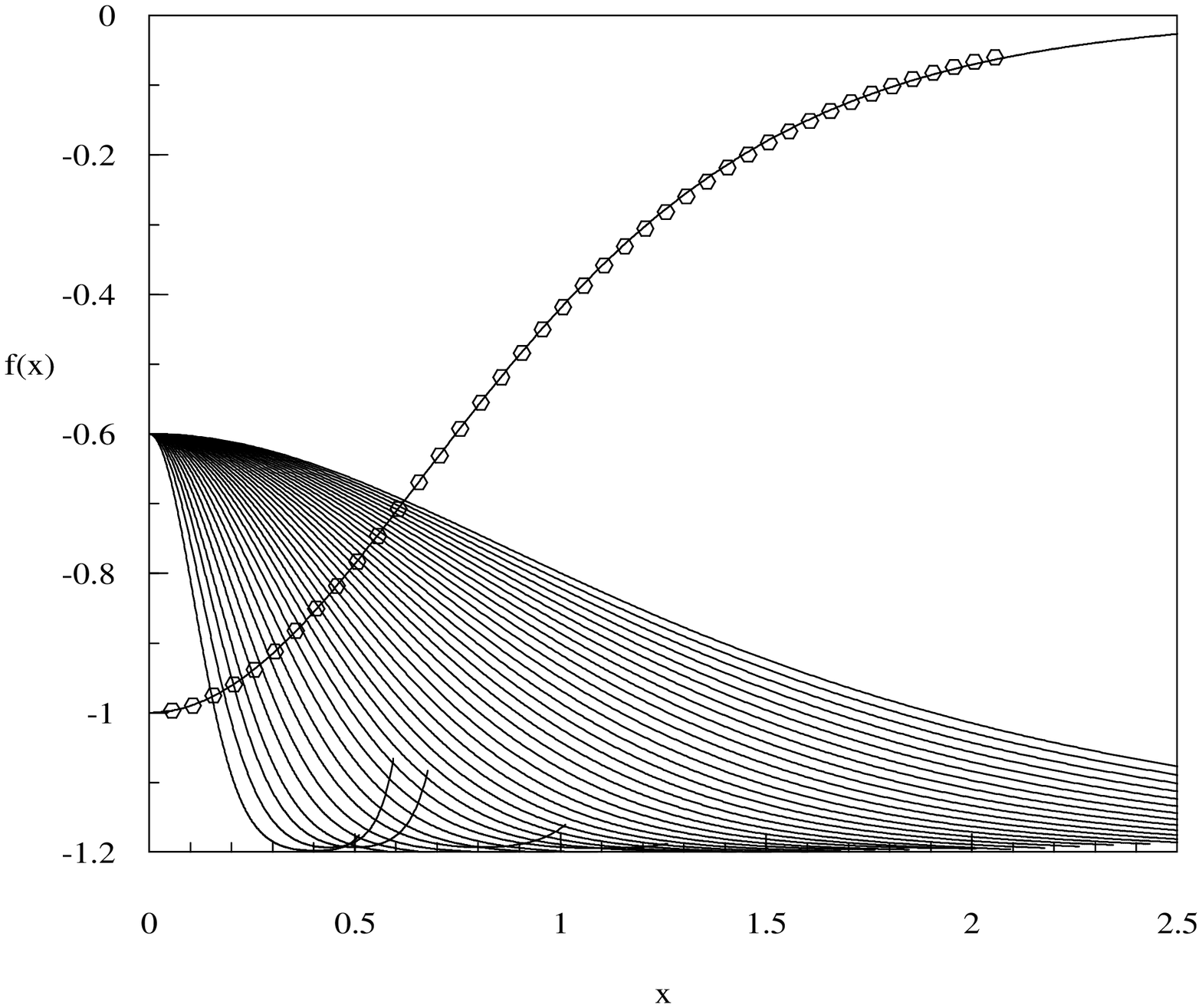,width=4in,height=3in}
\caption{Constructive inversion of the energy trajectory $F(v)$ for
 the sech-squared potential $f(x) = -{\rm sech}^{2}(x).$  For $x \leq b = 0.1,$ the
 algorithm correctly generates the model $f(x) = -1 + x^2;$ for larger values of
 $x,$ in steps of size $h = 0.05,$ the hexagons indicate the reconstructed values
 for the potential $f(x),$ shown exactly as a smooth curve. The unnormalized wave functions are also shown.}
\label{fig6}
\end{figure}
 The parameters used are exactly the same as for the first problem described above.  The time taken to perform the inversions was less than $20$ seconds if we discount, in the case of the exponential potential, the extra time taken to compute $F(v)$ itself.   
\section{Kinetic potentials} 
Geometry is involved with this problem because we deal with a family of operators depending on a continuous parameter $v.$  This immediately leads to a family of spectral manifolds, and, more particularly, to the consideration of smooth transformations of potentials, and to the transformations which they in turn induce on the spectral manifolds.   This is the environment in which we are able to construct the following functional inversion sequence:

$$f^{[n+1]} = \bar{f}\circ\bar{f}^{[n]^{-1}}\circ f^{[n]} \equiv \bar{f}\circ K^{[n]}.\eqno{(6.1)}$$

\noindent A {\it kinetic potential} is the constrained mean value of the potential shape $\bar{f}(s) = \langle f\rangle,$ where the corresponding mean kinetic energy $s = \langle -\Delta\rangle$ is held constant.  It turns out that kinetic potentials may be obtained from the corresponding energy trajectory $F$ by what is essentially a Legendre transformation \cite{gel} $\bar{f} \leftrightarrow F$ given \cite{env3} by

$$\bar{f}(s) = F'(v),\quad s = F(v)-vF'(v),$$
and
$$F(v)/v = \bar{f}(s) - s\bar{f}'(s),\quad 1/v = - \bar{f}'(s).\eqno{(6.2)}$$

\noindent These transformations are well defined because of the definite convexities of $F$ and $\bar{f};$  they complete the definition of the inversion sequence (1.2), up to the choice of a starting seed potential $f^{[0]}(x).$  They differ from  Legendre transformations only because of our choice of signs.  The choice has been made so that the eigenvalue can be written (exactly) in the semi-classical forms

$$E = F(v) = \min_{s > 0}\left\{s + v\bar{f}(s)\right\} = \min_{x > 0}\left\{K^{[f]}(x) + vf(x)\right\}\eqno{(6.3)}$$

\noindent where the kinetic- and potential-energy terms have the `usual' signs.  
\section{Envelope theory}
The term `kinetic potential' is short for `minimum mean iso-kinetic potential'.  If the Hamiltonian is $H = -\Delta + vf(x),$ where $f(x)$ is potential {\it shape,} and ${\cal D}(H) \subset L^{2}(\Re)$ is the domain of $H,$ then the ground-state kinetic potential $\bar{f}(s) = \bar{f}_{0}(s)$ is defined \cite{env2,env3} by the expression

$$\bar{f}(s) = \inf_{{{\scriptstyle \psi \in {\cal D}(H)} \atop {\scriptstyle (\psi,\psi) = 1}} \atop {\scriptstyle (\psi, -\Delta\psi) = s}} (\psi, f\psi).\eqno{(7.1)}$$

The extension of this definition to the higher discrete eigenvalues (for $v$ sufficiently large) is straightforward \cite{env3} but not explicitly needed in the present paper. The idea is that the min-max computation of the discrete eigenvalues is carried out in two stages: in the first stage (7.1) the mean potential shape is found for each fixed value of the mean kinetic energy $s;$ in the second and final stage we minimize over $s.$  Thus we have arrive at the semi-classical expression which is the first equality of Eq.(1.4).  It is well known that $F(v)$ is concave ($F''(v) < 0$) and it follows immediately that $\bar{f}(s)$ is convex.  More particularly, we have \cite{inv1} 

$$F''(v)\bar{f}''(s) = -{1 \over {v^{3}}}.\eqno{(7.2)}$$

Thus, although kinetic potentials are {\it defined} by (7.1), the transformations (6.2) may be used in practice to go back and forth between $F$ and $\bar{f}.$

Kinetic potentials have been used to study smooth transformations of potentials and also linear combinations.  The present work is an application of the first kind. Our goal is to devise a method of searching for a transformation $g,$ which would convert the initial seed potential $f^{[0]}(x)$ into the (unknown) goal $f(x) = g(f^{[0]}).$  We shall summarize briefly how one proceeds in the forward direction, to approximate $F,$ if we know $f(x).$  The $K$ functions are then introduced, by a change of variable, so that the potential $f(x)$ is exposed and can be extracted in a sequential inversion process.

In the forward direction we assume that the lowest eigenvalue $F^{[0]}(v)$ of $H^{[0]} = -\Delta + vf^{[0]}(x)$ is known for all $v > 0$ and we assume that $f(x)$ is given; hence, since the potentials are symmetric and monotone for $x > 0,$ we have defined the transformation function $g.$  `Tangential potentials' to $g(f^{[0]})$ have the form $a + bf^{[0]}(x),$ where the coefficients $a(t)$ and $b(t)$ depend on the point of contact $x = t$ of the tangential potential to the graph of $f(x).$  Each one of these tangential potentials generates an energy trajectory of the form ${\cal F}(v) = av + F^{[0]}(bv),$ and the {\it envelope} of this family (with respect to $t$) forms an approximation $F^{A}(v)$ to $F(v).$  If the transformation $g$ has definite convexity, then $F^{A}(v)$ will be either an upper or lower bound to $F(v).$ It turns out \cite{env3} that all the calculations implied by this envelope approximation can be summarized nicely by kinetic potentials. Thus the whole procedure just described corresponds exactly to the expression:
\[
\bar{f} \approx \bar{f}^{A} = g\circ\bar{f}^{[0]},\eqno{(7.3)}
\]
\noindent with $\approx$ being replaced by an inequality in case $g$ has definite convexity. Once we have an approximation $\bar{f}^{A},$ we immediately recover the corresponding energy trajectory $F^{A}$ from the general minimization formula (6.3).  

The formulation that reveals the potential shape is obtained when we use $x$ instead of $s$ as the minimization parameter.  We achieve this by the following general definition of $x$ and of the $K$ function associated with $f:$
$$f(x) = \bar{f}(s),\quad K^{[f]}(x) = \bar{f}^{-1}(f(x)).\eqno{(7.4)}$$
\noindent The monotonicity of $f(x)$ and of $\bar{f}$ guarantee that $x$ and $K$ are well defined. Since $\bar{f}^{-1}(f)$ is a convex function of $f,$ the second equality in (1.4) immediately follows \cite{env2}.  In terms of $K$ the envelope approximation (2.3) becomes simply
$$K^{[f]} \approx K^{\left[f^{[0]}\right]}.\eqno{(7.5)}$$
\noindent Thus the envelope approximation involves the use of an approximate $K$ function that no longer depends on $f,$ and there is now the possibility that we can invert (1.4) to extract an approximation for the potential shape.

We end this summary by listing some specific results that we shall need.  First of all, the kinetic potentials and $K$ functions obey \cite{env1,env2} the following elementary shift and scaling laws:
$$f(x) \rightarrow a + bf(x/t) \Rightarrow  \left\{\bar{f}(s) \rightarrow a + b\bar{f}(st^{2}),\quad K^{[f]}(x) \rightarrow {1 \over {t^2}}K^{[f]}\left({x \over t}\right)\right\}.\eqno{(7.6)}$$
\noindent Pure power potentials are important examples which have the following formulas:
$$f(x) = |x|^{q} \Rightarrow \left\{\bar{f}(s) = \left ({P \over {s^{1 \over 2}}}\right )^{q}, \quad
K(x) = \left ({P \over x}\right )^{2}\right \},\eqno{(7.7)}$$
\noindent where, if the bottom of the spectrum of $-\Delta + |x|^{q}$ is $E(q),$ then the $P$ numbers are given \cite{env2} by the following expressions with $n = 0:$
$$P_{n}(q) = \left\vert E_{n}(q)\right\vert^{{(2+q)} \over {2q}}\left[{2 \over {2+q}}\right]^{1 \over q}\left[{{|q|} \over {2+q}}\right]^{1 \over 2}, \quad q \neq 0.\eqno{(7.8)}$$
We have allowed for $q < 0$ and for higher eigenvalues since the formulas are essentially the same.  The $P_{n}(q)$ as functions of $q$ are interesting in themselves \cite{env2}: they have been proved to be monotone increasing, they are probably concave, and $P_{n}(0)$ corresponds exactly to the $\log$ potential.  By contrast the $E_{n}(q)$ are not so smooth: for example, they have infinite slopes at $q = 0.$  But this is another story.  An important observation is that the $K$ functions for the pure powers are {\it all} of the form $(P(q)/x)^{2}$ and they are invariant with respect to both potential shifts and multipliers: thus $a + b|x|^{q}$ has the same $K$ function as does $|x|^{q}.$  For the harmonic oscillator $P_n(2) = (n+{1 \over 2})^{2}, \quad n = 0,1,2,\dots.$ Other specific examples may be found in the references cited. 

The last formulas we shall need are those for the ground state of the sech-squared potential:
$$f(x) = -{\rm sech}^{2}(x) \Rightarrow \left\{\bar{f}(s) = -{{2s} \over {(s + s^2)^{1 \over 2} + s}}, \quad K(x) = {\rm sinh}^{-2}(2x)\right\}.\eqno{(7.9)}$$
 \section{Functional inversion} 
The inversion sequence (6.1) is based on the following idea. The goal is to find a transformation $g$ so 
that $f = g\circ f^{[0]}.$  We choose a seed $f^{[0]},$ but, of course, $f$ is unknown.  In so far as the envelope approximation with $f^{[0]}$ as a basis is `good', then an approximation $g^{[1]}$ for $g$ would be given by $\bar{f} = g^{[1]}\circ\bar{f}^{[0]}.$ Thus we have

 $$g \approx g^{[1]} = \bar{f}\circ\bar{f}^{[0]^{-1}}.\eqno{(8.1)}$$

Applying this approximate transformation to the seed we find:

$$f \approx f^{[1]} = g^{[1]}\circ f^{[0]} = \bar{f}\circ\bar{f}^{[0]^{-1}}\circ f^{[0]} = \bar{f}\circ K^{[0]}.\eqno{(8.2)}$$

We now use $f^{[1]}$ as the basis for another envelope approximation, and, by repetition, we have the ansatz (1.2), that is to say

$$f^{[n+1]} = \bar{f}\circ\bar{f}^{[n]^{-1}}\circ f^{[n]} = \bar{f}\circ K^{[n]}.\eqno{(8.3)}$$

A useful practical device is to invert the second expression for $F$ given in (1.4) to obtain

$$K^{[f]}(x) = \max_{v > 0}\left\{F(v) - vf(x)\right\}.\eqno{(8.4)}$$

The concavity of $F(v)$ explains the $\max$ in this inversion, which, as it stands, is exact.  In a situation where $f$ is unknown, we have $f$ on both sides and nothing can be done with this formal result.  However, in the inversion sequence which we are considering, (3.4) is extremely useful.  If we re-write (3.4) for stage [n] of the inversion sequence it becomes:

$$K^{[n]}(x) = \max_{v > 0}\left\{F^{[n]}(v) - vf^{[n]}(x)\right\}.\eqno{(8.5)}$$

\noindent In this application, the current potential shape $f^{[n]}$ and consequently $F^{[n]}(v)$ can be found (by shooting methods) for each value of $v.$  The minimization can then be performed even without differentiation (for example, by using a Fibonacci search) and this is a much more effective method for $K^{[n]} = \bar{f}^{[n]^{-1}}\circ f^{[n]}$ than finding $\bar{f}^{[n]}(s),$ finding the functional inverse, and applying the result to $f^{[n]}.$   
\section{Functional inversion for pure powers} 
We now treat the case of pure-power potentials given by

$$f(x) = A + B|x|^{q}, \quad q > 0,\eqno{(9.1)}$$

\noindent where  $A$ and $B > 0$ are arbitrary and fixed.  We shall prove that, starting from another pure power as a seed, the inversion sequence converges in just two steps. The exact energy trajectory $F(v)$ for the potential (9.1) is assumed known.  Hence, so is the exact kinetic potential given by (7.7) and the general scaling rule (7.6), that is to say  

$$\bar{f}(s) = A + B\left ({P(q) \over {s^{1 \over 2}}}\right )^{q}.\eqno{(9.2)}$$

We now suppose that a pure power is also used as a seed, thus we have

$$f^{[0]}(x) = a + b|x|^{p}\quad \Rightarrow \quad K^{[0]}(x) = \left({{P(p)} \over x}\right)^{2},\eqno{(9.3)}$$

where the parameters $a,\quad b > 0, \quad p > 0$ are arbitrary and fixed.  The first step of the inversion (6.3) therefore yields

$$f^{[1]}(x) = \left(\bar{f}\circ K^{[0]}\right)(x) = A + B\left({{P(q)|x|} \over {P(p)}}\right)^{q}.\eqno{(9.4)}$$

The approximate potential $f^{[1]}(x)$ now has the correct $x$ power dependence but has the wrong multiplying factor.  Because of the invariance of the $K$ functions to multipliers, this error is completely corrected at the next step, yielding:

$$K^{[1]}(x) =  \left({{P(q)} \over x}\right)^{2}\quad\Rightarrow\quad f^{[2]}(x) = \left(\bar{f}\circ K^{[1]}\right)(x) = A + B|x|^q.\eqno{(9.5)}$$

This establishes our claim that power potentials are inverted without error in exactly two steps. 

The implications of this result are a little wider than one might first suspect.  If the potential that is being reconstructed has the asymptotic form of a pure power for small or large $x,$ say, then we know that the inversion sequence will very quickly produce an accurate approximation for that part of the potential shape.  More generally, since the first step of the inversion process involves the construction of $K^{[0]},$ the general invariance property $K^{[a+bf]} = K^{[f]}$ given in (7.6) means that the seed potential $f^{[0]}$ may be chosen without special consideration to gross features of $f$ already arrived at by other methods. For example, the area (if the potential has area), or the starting value $f(0)$ need not be incorporated in $f^{[0]},$ say, by adjusting $a$ and $b.$     
\section{A more general example}
We consider the problem of reconstructing the sech-squared potential $f(x) = -{\rm sech}^{2}(x).$  We assume that the corresponding exact energy trajectory $F(v)$ and, consequently, the kinetic potential $\bar{f}(s)$ are known.  Thus \cite{env2} from the potential shape $f(x) = -{\rm sech}^{2}(x)$ we have:
$$\left\{F(v) = -\left[\left(v+{1 \over 4}\right)^{1 \over 2} - {1 \over 2}\right]^{2},\quad\bar{f}(s) = -{{2s} \over {(s + s^2)^{1 \over 2} + s}}\right\}.\eqno{(10.1)}$$.
The seed is essentially $x^2,$ but we use a scaled version of this for the purpose of illustration in Fig.(7).  Thus we have 
$$f^{[0]} = -1 + {{x^2} \over 20}\quad\Rightarrow\quad K^{[0]}(x) = {1 \over {4x^{2}}}\eqno{(10.2)}$$
This potential generates the exact eigenvalue
$$F^{[0]}(v) = -v + \left({v \over {20}}\right)^{1 \over 2}\eqno{(10.3)}$$,
which, like the potential itself, is very different from that of the goal.  After the first iteration we obtain
$$f^{[1]}(x) = \bar{f}\left(K^{[0]}(x)\right) = -{2 \over {1 + (1+4x^{2})^{1 \over 2}}}.\eqno{(10.4)}$$
A graph of this potential is shown as $f1$ in Fig.(7). In order to continue analytically we would need to solve the problem with Hamiltonian $H^{[1]} = -\Delta + vf^{[1]}(x)$ exactly to find an expression for $F^{[1]}(v).$  We know no way of doing this. However, it can be done numerically, with the aid of the inversion formula (8.5) for $K.$  The first 5 iterations shown in Fig.(7) suggest convergence of the functional sequence.
\begin{figure}
\psfig{file=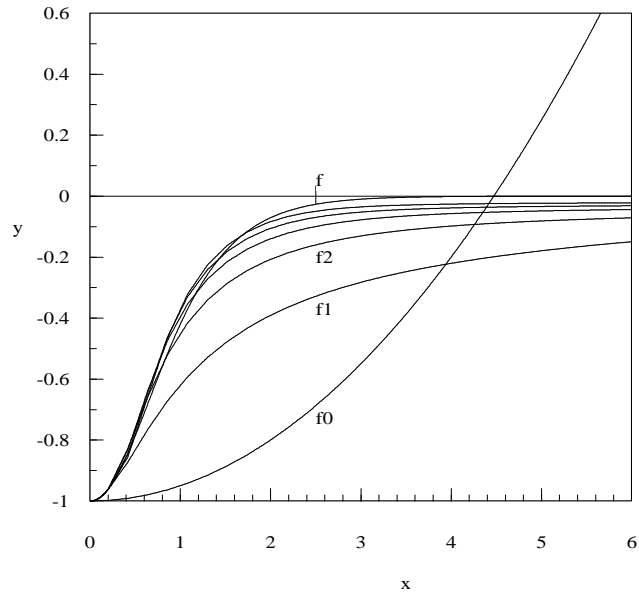,width=4in,height=4in}
\caption{The energy trajectory $F$ for the sech-squared potential $f(x) = -{\rm sech}^{2}(x)$ is approximately inverted starting from the seed $f^{[0]}(x) = -1 + x^{2}/20.$  The first step can be completed analytically yielding $f1 = f^{[1]}(x) = -2/\{1 + \sqrt{1 + 4x^{2}}\}.$ Four more steps $\{fk = f^{[k]}\}_{k=2}^{5}$ of the inversion sequence approaching $f$ are performed numerically.}
\label{fig7}
\end{figure}

\section{Conclusion}
We have discussed two methods that may be used to reconstruct the potential $f(r)$ in a Schr\"odinger operator $H = -\Delta +vf(r)$ if an eigenvalue curve $E = F_n(v)$ is known as a function of the coupling $v.$  Such functions as $F(v)$ are met, for example, as  one-body approximations to certain $N$-body systems, such as atoms. In addition to the  inversion of physical data to infer the structure of the underlying system, the situation also presents a rather fascinating mathematical problem.  Work  is under way extending these results and methods to a wider class of problems, including those with singular potentials.

\section*{Acknowledgements}
Partial financial support of this work under Grant No. GP3438 from the
Natural Sciences and Engineering Research Council of Canada, and the hospitality of
the organizers of the 13th Regional Conference on Mathematical Physics, Antalya, Turkey, are gratefully acknowledged.


\end{document}